 \documentstyle[twoside,fleqn,espcrc2,epsfig]{article}

\newcommand{\AmS}{{\protect\the\textfont2
  A\kern-.1667em\lower.5ex\hbox{M}\kern-.125emS}}

\title{Heavy Nuclei, from RHIC to the Cosmos}

\author{Spencer R. Klein\address[LBNL]{Lawrence Berkeley National Laboratory, 
Berkeley, CA, 94720, USA}}

\begin{document}

\begin{abstract}
Ultra-relativistic heavy ion collisions produce a high-temperature,
thermalized system that may mimic the conditions present shortly after
the big bang.  This writeup will given an overview of early results
from the Relativistic Heavy Ion Collider (RHIC), and discuss what we
have learned about hot, strongly interacting nuclear systems.  The
thermal and chemical composition of the system will be discussed,
along with observables that are sensitive to the early evolution of
the system. I will also discuss the implications of the RHIC results
for cosmic ray air showers.
\vspace{1pc}
\end{abstract}

\maketitle

\section{Introduction}

The Relativistic Heavy Ion Collider (RHIC) at Brookhaven National
Laboratory (BNL) collides ultra-relativistic ions at energies up to
200 GeV per nucleon.  The nucleon-nucleon reactions are energetic
enough that perturbative QCD is expected to be able to describe much
of the collision dynamics.

The goal of ultra-relativistic heavy ion collisions is to study the
properties of matter at extremely high temperatures and/or densities,
with an eye to mimicing the conditions present in the very early
universe, $\approx$ 10$\mu$s after the big bang.  A specific goal is
to look for the Quark-Gluon Plasma, a state of matter whereby the
protons and neutrons in a nucleus 'dissolve', producing a gas of free
quarks and gluons.

These collisions may also be similar to those produced when heavy-ion
cosmic rays hit the atmosphere.  In the target frame, RHIC projectile
gold nuclei have a total energy of 4.3 PeV (20 TeV per nucleon),
energetically reaching the knee of the cosmic ray spectrum.

Relativistic heavy ion collisions were initially studied at the
Berkeley Bevatron, SIS and the Dubna Nuclotron\cite{lowenergyrev}.
More recently, there have been higher energy studies at the BNL AGS
and the CERN SPS.  The SPS data is often used as a lower-energy
comparison point; the SPS collided lead on lead, at a center of mass
energy of 17 GeV per nucleon. The earlier studies found several
interesting phenomena. These include:

Anisotropic flow: Heavy ion collisions may be described at least
partly in terms of fluid dynamics; the system shows
fluidlike behavior\cite{flowrev}.

Strangeness enhancement: Production of strange particles is several
times larger than would be expected from superimposed $pp$
collisions\cite{sigrev}\cite{spiros}\cite{arizona}.

J/$\psi$ suppression: Production of J/$\psi$ particles is suppressed
compared to the production of Drell-Yan
dileptons\cite{sigrev}\cite{jpsirev}.

The latter two observations have been proposed as signatures of the
Quark Gluon Plasma.  However, both phenomena might be due to normal
hadronic interactions, with the additional strangeness produced in
secondary reactions among the produced hadrons, and the J/$\psi$
suppression due to interactions with the initial state nucleons and
the other hadrons produced in the collision.

In low energy heavy ion collisions, the interacting baryons stop when
the nuclei collide.  As the collision energy increases, the nuclei
gradually become transparent, the baryons retain some of their initial
momentum, and the net baryon density of the produced system drops.  At
RHIC, the net baryon density at mid-rapidity (near the system center
of mass) should be near zero.

This writeup will discuss heavy ion collisions at RHIC, starting with
observables that probe thermal freezeout, such as the global event
characteristics, system size, particle spectra, and non-isotropic
flow.  Next, the composition at chemical freezeout will be discussed,
followed by signatures of the early evolution, focusing on high $p_T$
particles and charm production.

\section{RHIC}

RHIC is a 3.8 km circumference double-ring accelerator which can
collide gold ions at center of mass energies of up to 200 GeV/nucleon
at a luminosity of $2\times10^{26}/$cm$^2/$s, corresponding to about
1,500 hadronic collisions/sec.  RHIC can also accelerate lighter
ions. The maximum energy per nucleon depends on the charge to mass
ratio; for protons, the maximum center of mass energy is 500 GeV.  It
also collides polarized protons, to study the spin structure of the
nucleon.  The luminosity depends on the species; for protons the
luminosity can reach $1.4\times10^{31}$/cm$^2$/s, or about 700,000
hadronic interactions/sec.

In the year 2000, RHIC collided gold nuclei at an energy of 130
GeV/nucleon.  Most of the results presented here are from this run.
In 2001/2, RHIC collided gold nuclei at 200 GeV/nucleon, briefly
reaching the design luminosity, and collided polarized protons, with
up to 25\% polarization.  The long term program will include studies
with lighter ions, gold-gold collisions at lower energies, and
deuterium-gold and/or proton-gold collisions.

RHIC is instrumented with two large detectors, STAR and PHENIX, and
two smaller experiments, BRAHMS and PHOBOS. A third small experiment,
$pp2pp$, studies proton-proton elastic scattering\cite{pp2pp}.  The
collaborations have very different strategies for studying ion
collisions.

PHENIX is designed to look for relatively rare observables that are
sensitive to the early phases of the collision, such as charmed
hadrons, $J/\psi$ and direct photons\cite{PHENIX}.  The detectors are
optimized for particle identification, especially leptons and photons.
PHENIX has a two-armed central spectrometer; each arm is instrumented
with charged particle tracking, time-of-flight (TOF), a ring imaging
Cherenkov counter (RICH), and electromagnetic calorimetry.  Each arm
covers a solid angle of 135 degrees in azimuth by 0.3 in
pseudorapidity, where the pseudorapity $\eta=-\ln[\tan(\theta/2)]$,
with $\theta$ the particle angle with respect to the beam axis.  The
center of mass is at $\eta=0$.  Forward and backward muon detectors
cover $1.2< |\eta|< 2.2$ for muons with momentum $p>2$ GeV/c.
Specialized triggers and a high rate DAQ system will collect large
samples of the selected rare probes.

The Solenoidal Tracker at RHIC (STAR) is optimized to study hadrons
over a very large solid angle, including multi-particle correlations,
and measure global event characteristics\cite{STAR}.

STAR tracks charged particles with $\eta|<1.5$ in a large time
projection chamber (TPC) in a 5 kG solenoidal magnetic field.  A
silicon vertex detector covering $|\eta|<1$ and two forward TPCs
covering $2.5<|\eta|<4.0$ complete the tracking system.  Strange
particles like $K_S$, $\Lambda$, $\Xi$ and $\Omega$ are detected by
reconstructing secondary vertices.  Energy loss in the TPCs and SVT
and small TOF and RICH systems provide particle identification, along
with an electromagnetic calorimeter.  STAR records a great deal of
information on each event, but can only record data from selected
events at rate slower than PHENIX.

PHOBOS records charged and neutral particle production over most of
phase space, up to $|\eta|<5.4$, to search for anomalous event
shapes\cite{PHOBOS}.  It has two small charged particle spectrometers
with TOF systems for particle identification.

BRAHMS is composed of precision central and forward spectrometers with
tracking and particle identification, along with counters to measure
charged multiplicity\cite{BRAHMS}.

The 4 experiments include identical zero degree calorimeters (ZDCs) to
measure forward neutrons from nuclear fragmentation\cite{ZDCs}.  The
ZDCs are intended to provide a common method for luminosity and
centrality (impact parameter) measurements, in order to facilitate
comparisons between the four experiments.  

\section{Ultra-Peripheral Collisions}

Before discussing central collisions, it is interesting to consider
ultra-peripheral collisions (UPCs), interactions at large impact
parameters $b$ (minimum ion-ion separation) where only photonuclear
and two-photon interactions are possible. UPCs can probe a wide
variety of physics\cite{reviews}, ranging from electrodynamics in very
strong electromagnetic fields to meson spectroscopy to measurements of
gluon shadowing in heavy nuclei to tests of quantum
mechanics\cite{rhointerference}.  Photoproduction of heavy
hadrons\cite{hq} and quarkonium\cite{strikman} is sensitive to the
gluon density in the nucleus, and hence to gluon shadowing.

At RHIC, mutual nuclear excitation (including both Coulombic and
hadronic interactions) is used as a luminosity
monitor\cite{luminosity}.  The process has a large cross section,
about 11 barns, and small backgrounds.

\begin{figure}
\setlength{\epsfxsize=2.8 in}
\setlength{\epsfysize=1.8 in}
\centerline{\epsffile{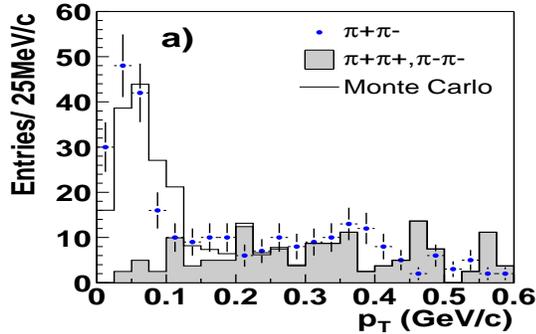}}
\caption[]{$p_T$ spectrum of 2 track events observed at 130 GeV in
STAR.  The peak at low $p_T$ is characteristic of coherent coupling to
both nuclei, as expected for coherent $\rho^0$
photoproduction\cite{UPCrho}.}
\label{fig:pcpt}
\end{figure}

One easily observable UPC process is exclusive coherent
photoproduction of vector mesons, $Au\!+\! Au\! \rightarrow\! Au\! +\!
Au\! + \rho^0$. These events are characterized by an almost empty
detector, containing only two tracks, with a total event transverse
momentum $p_T<2\hbar/R_A\approx 100$ MeV/c.  The low $p_T$,
characteristic of the coherent photon emission and scattering, is a
distinctive signature, as data from STAR shows in Fig. \ref{fig:pcpt}.
At 130 GeV, STAR measures $\sigma(Au\!+\!Au\!\rightarrow\!Au\!+\!Au\!+\!
\rho^0) = 460 \pm 220\pm 110$ mb\cite{UPCrho}, in agreement with
theoretical predictions\cite{rhorates}.  The ratio of $\rho^0$ to
direct $\pi^+\pi^-$ production is consistent with that measured in
$\gamma p$ interactions.

\section{Hadronic Collisions}

\begin{figure}
\setlength{\epsfxsize=2.8 in}
\setlength{\epsfysize=2.0 in}
\centerline{\epsffile{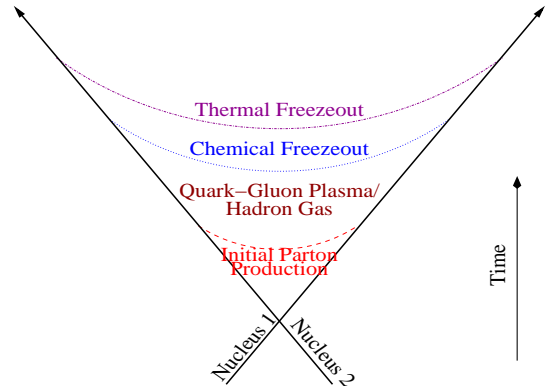}}
\caption[]{Schematic view of a heavy ion interaction, showing the
different stages of the reaction.}
\label{initschematic}
\end{figure}

Hadronic collisions occur in several stages, as is shown in
Fig. \ref{initschematic}.  The nucleons collide, and their partons
interact.  The produced particles interact and form hadrons
(hadronize).  As the interactions continue and the number of particles
grows, the system expands and cools, When the average particle energy
is low enough, inelastic hadron production stops, a transition known
as chemical freezeout.  Slightly later, the interparticle separation
is large enough that even elastic interactions cease; this is thermal
freezeout.

The key question in this picture is whether the produced particles
interact as hadrons (i.e. a hadron gas) or as partons (i.e., a
quark-gluon plasma). Partons produced in the initial interactions may
remain free for long enough to interact with each other and
equilibrate, forming a quark-gluon plasma.  Or, they might immediately
form hadrons (hadronize), and the interacting system will be a hadron
gas.  Or, they could initial interact as a quark-gluon plasma, and,
then, as the system cools, hadronize to form a hadron gas.

Most of the theoretical guidance regarding the quark-gluon plasma
comes from lattice gauge theory (LGT).  Recent LGT calculations
indicate that the phase transition between hadron gas and quark-gluon
plasma, if it occurs, is weak (at least second order), and occurs at a
temperature of 150-200 MeV and an energy density $\epsilon_c\approx 1$
GeV/fm$^3$\cite{lattice}.  This calculation is for an infinite medium
with an infinite lifetime; edge effects and formation time are not
considered.  Although the expected system lifetime is only
$\approx10^{-23}$s, calculations indicate that equilibration occurs
quickly, so a clear phase change is possible.

\begin{figure}
\setlength{\epsfxsize=2.7 in}
\setlength{\epsfysize=2.7 in}
\centerline{\epsffile{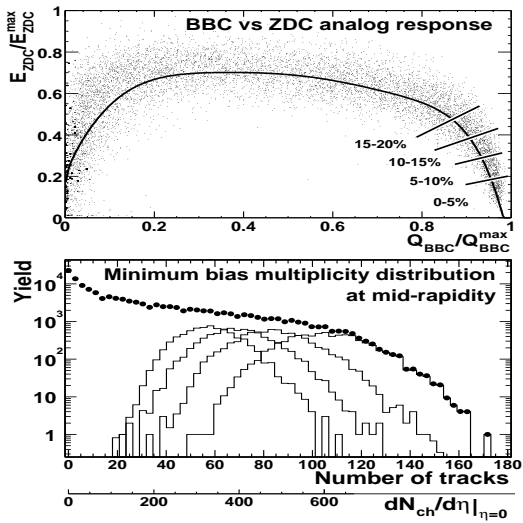}}
\caption[]{(a)Relationship between forward neutrons (ZDC energy) and
charged multiplicity (the charge in the beam-beam
counters, $Q_{BBC}$), as measured by the PHENIX collaboration at 130
GeV for 4 impact parameter bins.  (b) The overall charged particle
multiplicity, $d\sigma/dN_{ch}$ (solid dots), with calculations of the
multiplicity distribution for the same 4 impact parameter bins.  From
Ref. \cite{phenixmult}.}
\label{fig:phenixmult}
\end{figure}

Although it is a key parameter in heavy ion collisions, the impact
parameter $b$ is not directly observable.  We use two classes of
observables to infer the impact parameter.  The first is the number of
forward (zero-degree) neutrons.  These neutrons come from the
non-interacting part of the nucleus.  Enough energy propagates from
the initial collision to dissociate the non-interacting part of the
nuclei into neutrons, protons and small nuclear fragments. The other
observables are sensitive to the number of interacting nucleons.
Examples are the charged particle multiplicity or transverse energy.
A model is necessary to relate these observables to the impact
parameter.  To avoid systematic uncertainties, events are often sorted
by centrality ({\it i.e.} by charged multiplicity), and divided into
classes, such as the 10\% most central (those with the smallest impact
parameter).

Figure \ref{fig:phenixmult} shows the relationship between the number
of forward neutrons (measured in the ZDCs) and the charged
multiplicity\cite{phenixmult}.  The charged multiplicity rises
continually as the impact parameter decreases.  However, the number of
forward neutrons is largest at moderate impact parameters.  In very
central collisions, most of the nucleus interacts, leaving few
remnant neutrons, while in very peripheral collisions, some of the
nucleus remains intact, reducing the number of forward neutrons,
producing the curve in Fig. \ref{fig:phenixmult}.  

\section{Thermal Freezeout}

\begin{figure}
\setlength{\epsfxsize=2.8 in}
\setlength{\epsfysize=2.4 in}
\centerline{\epsffile{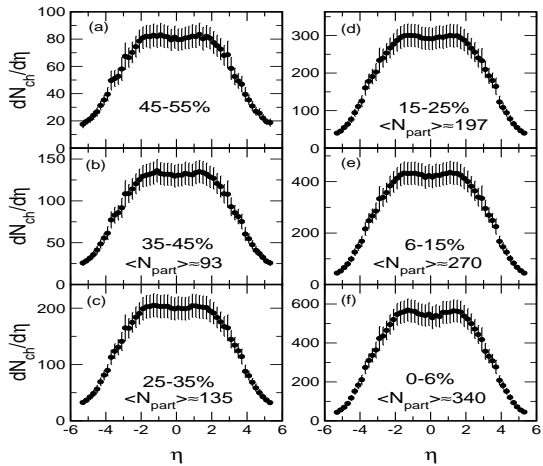}}
\caption[]{Charged multiplicity $dN/d\eta$ for different centrality
bins at 130 GeV. A flat plateau is visible at mid-rapidity. From the
PHOBOS collaboration\cite{phobosmult}.}
\label{phobosdndy}
\end{figure}

The particles present at thermal freezeout are those observed in the
RHIC detectors, and are relevant for comparison with models of heavy
ion collisions.  The charged particle multiplicity is shown as a
function of pseudorapidity $\eta$ in Fig. \ref{phobosdndy}.  The
multiplicity $dN/d\eta$ is roughly flat for $|\eta|<2$. This central
plateau shows that there is boost invariance.  Within this region, the
system appears invariant with respect to the longitudinal boost
(velocity); the expansion may be treated in 2 dimensions.

At 130 GeV, the maximum $dN/d\eta$ is about 570, rising to $650$ at
200 GeV.  This corresponds to total multiplicities of about
$4100\pm210$ and $4960\pm250$ respectively\cite{WitSB}.  These
multiplicities are considerably lower than most pre-RHIC
predictions\cite{eskola}, and seem to be best fit by models based on a
combination of hard interactions (calculated by perturbative QCD) and
soft interactions (extrapolated from lower energies).  Most popular
cosmic ray air shower codes predict considerably larger
multiplicities\cite{Engel}.

The $dN/dy$ per participant (nucleon involved in the collision) are
about 40\% higher than at lower energies, and also 50\% higher than in
$\overline p p$ collisions at comparable energies\cite{phobosmult}.
The multiplicity per participant rises smoothly as the number of
participants increases.

The PHENIX collaboration measured a transverse energy
$dE_T/d\eta\approx 578$ GeV per unit $\eta$ for the 2\% most central
collisions.  This is the energy released in the collision. In a
boost-invariant picture (supported by the existence of a central
plateau), particles emitted into a pseudorapidity region $\delta\eta$
cine from a region of longitudinal size $\delta\eta c\tau$, where
$\tau\approx\hbar/p_0$ is the time required for the initial
interactions to occur.  Here, $p_0$ is the energy scale for the
initial particle production.  The initial energy density depends on
this scale.  We will use a conservative $p_0=200$ MeV/c, so
$\tau\approx 1$ fm/c.  The initial volume is $\pi R_A^2\delta\eta\tau$
and the energy density $\epsilon$ is
\begin{equation}
\epsilon \approx {1\over \pi R_A^2\tau} {dE_T\over d\eta} 
\approx 4.5\ {\rm GeV/fm}^3.
\end{equation}
This is much larger than the 1 GeV/cm$^3$ that lattice gauge
calculations predict is required to form a QGP.  Although the time
scale is not completely fixed, it seems hard to stretch $\tau$ enough
to reduce $\epsilon$ below 1 GeV/cm$^3$.

The baryon:antibaryon ratio at freezeout is also of interest.  Some
baryons are initially present in the gold nuclei, while the rest are
produced via baryon-antibaryon pair production.  Antibaryons come only
from the latter source.  At 130 GeV, the $\overline p:p$ ratio is
$0.6\pm0.02\pm0.06$, rising to $0.73\pm0.03$ for
$\overline\Lambda:\Lambda$ and $0.82\pm 0.08$ for $\overline\Xi:\Xi$
where the first (usually only) error is statistical\cite{helen}.

These ratios are quite close to 1. The central region is nearly baryon
free.  Pair produced baryons outnumber initial state baryons by more
than 2:1.  However, the net baryon number is not zero, showing that
there is indeed substantial baryon stopping; many initial state
baryons are transported over 6 units of rapidity.

The size of the system at thermal freezeout has been measured with
2-particle interferometry (Hanbury-Brown Twiss interferometry), taking
advantage of the Bose statistics that increase the abundance of
particle pairs with momentum difference $\Delta p = p_1-p_2 <
\hbar/R$.  Here $R$ is the radius of the last elastic interaction.
The momentum difference vector is decomposed into longitudinal (along
the beam direction), side (transverse to the observer, and out (toward
the observer) components (the Bertsch-Pratt decomposition). The source
radius for a source with an assumed Gaussian is about 6 fermi in all 3
dimensions\cite{hbt}.  This is about twice the initial nuclear radius
(about 6.5 fermi, in a Woods-Saxon [almost hard sphere] density
distribution).  This source size is similar to that observed in much
lower energy collisions at the SPS; the lack of growth is a surprise.
It is also surprising that the source radii in all 3 dimensions are
similar; this indicates that the particles are emitted in a very short
time scale, {\it i.e.} that thermal freezeout occurs quite suddenly
over the entire nucleus.

\begin{figure}
\setlength{\epsfxsize=2.8 in}
\setlength{\epsfysize=1.4 in}
\centerline{\epsffile{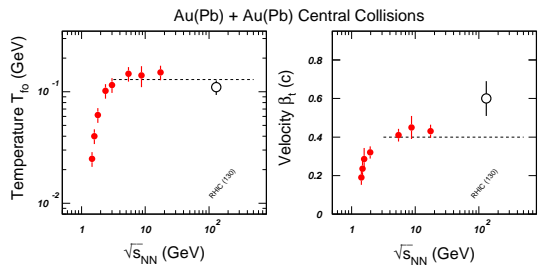}}
\caption[]{Temperatures $T$ and expansion velocity $\beta$ observed at
different collision energies. From Ref. \cite{kanetaxu}.}
\label{fig:thermal}
\end{figure}

Finally, we can compare the charged pion, proton and kaon spectra.
For $p_T < 2$ GeV/c (the non-perturbative region), all three spectra
are consistent with thermal emission.  However, the three species have
rather different apparent temperatures.  The different temperatures
may be due to a collective outward motion known as radial flow.  If
the expanding particles interact strongly, they tend to move outward
at the same velocities.  Then, the thermal fit temperature $T_{app}$
for a particle with mass $m$ is
\begin{equation}
T_{app} = T + m\beta^2
\label{eq:thermal}
\end{equation}
where $T$ is the actual temperature and $\beta\cdot c$ is the
collective expansion velocity.  The 3 species satisfy
Eq. \ref{eq:thermal} for $T=120$ MeV and $\beta=0.52$\cite{kanetaxu}.
The system expands outward at more than half the speed of light!  As
Fig. \ref{fig:thermal} shows, the temperature is slightly lower than
that measured at the SPS, but $\beta$ is significantly higher.  The
temperature is comparable to the transition temperature predicted by
LGT calculations. The very large $\beta$ is characteristic of
explosive expansion, with very high pressures and strong rescattering.

\begin{figure}
\setlength{\epsfxsize=2.8 in}
\setlength{\epsfysize=1.7 in}
\centerline{\epsffile{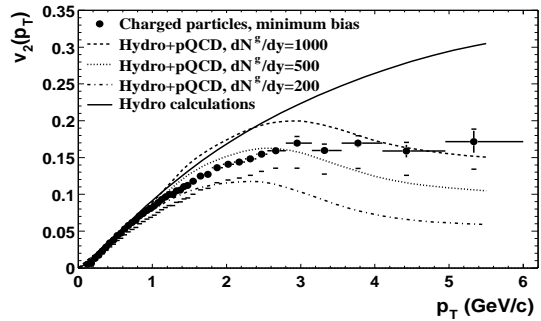}}
\caption[]{Elliptic flow ($v_2)$ as a function of $p_T$.  For $p_T< 2$
GeV/c, hydrodynamic models fit the data well.  At higher energies,
hydrodynamic models fail, as expected, and a parton description may be
more appropriate.  The data indicate that very high parton (mostly
gluon) densities are required to produce the observed flow.  From the
STAR collaboration \cite{STARhighptflow}.}
\label{starflow}
\end{figure}

Anisotropic flow is another observable.  In a non-central collision,
the reaction zone is elliptical.  Pressure converts this spatial
asymmetry into a particle density/momentum anisotropy which is usually
parameterized as
\begin{equation}
{dN\over d\phi} = 1 + 2v_2\cos{(2\phi)}
\end{equation}
where $\phi$ the angle between the particle and the reaction plane
(impact parameter vector), and $v_2$ is the elliptic flow.  A large
$v_2$ indicates high pressures and early equilibration\cite{flow}.
Figure \ref{starflow} shows the elliptic flow as a function of
centrality, here given in terms of $N_{ch}/N_{max}$, where $N_{ch}$ is
the charged particle multiplicity relative to the maximum multiplicity
$N_{max}$.  The flow is large, and is very close to the predictions of
hydrodynamic models that treat the system as a fluid.  Flow has been
studied for identified pions, protons and kaons ($K^\pm$ and $K_s$),
and $\Lambda$. The $p_T$ dependence of these species flow matches the
predictions of hydrodynamic models quite well\cite{STARidflow}.  This
fluidlike behavior is another indication of a strongly interacting
system.

\section{Chemical Freezeout}

\begin{figure}
\setlength{\epsfxsize=2.7 in}
\setlength{\epsfysize=2.4 in}
\centerline{\epsffile{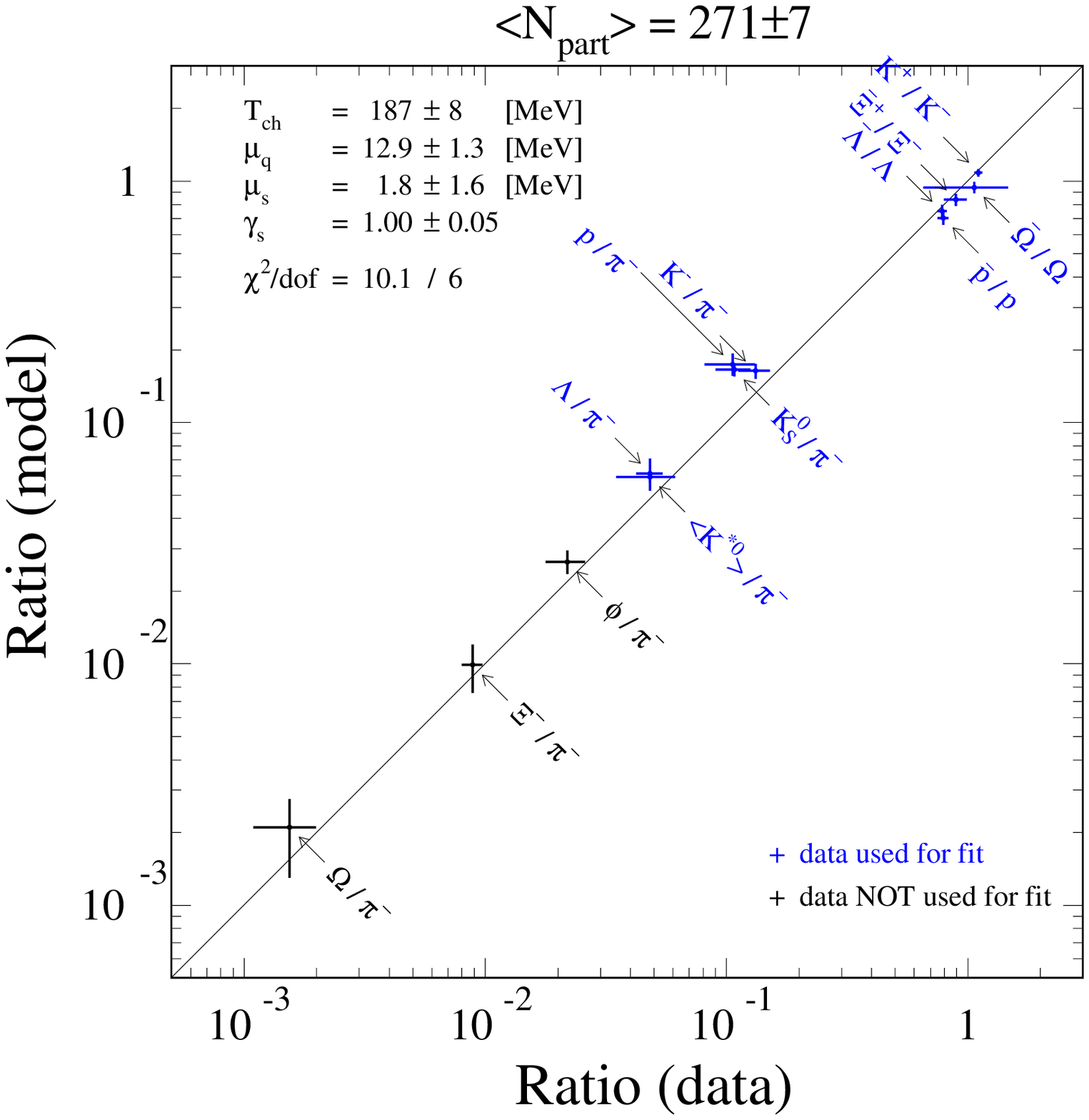}}
\caption[]{Measured particle multiplicity ratios vs.  the results of a
chemical fit model for the 6\% most central gold-gold collisions at
130 GeV.  The model axis errors are the uncertainties in the fit
output\cite{masashinu}.}
\label{fig:chemical}
\end{figure}

A key observable of chemical freezeout are the abundances of various
particles.  If the system is in chemical equilibrium, the abundances
should scale as $\exp{(-[\sqrt{m^2+p_T^2}+\mu]/kT)}$, where $m$ is the
particle mass, $\mu$ is the chemical potential of the particle (due to
it's baryon and strangeness content), $k$ is Boltzmann's constant, and
$T$ is the temperature\cite{kanetaxu}.  Figure \ref{fig:chemical}
compares the ratios of a number of different particles, compared with
the thermal model predictions.  The fit finds temperature $T= 187\pm
8$ MeV, baryochemical potential $\mu_b=39\pm 4$ MeV, strange chemical
potential $\mu_s = 1.8\pm1.6$ MeV, and strangeness suppression factor
$\gamma_s=1.00\pm0.05$.  This is hotter than at thermal freezeout,
which is a later, cooler stage in the evolution of the
system. The $\mu_b$ is much less than the proton mass, showing
quantitatively that the central region is effectively net baryon free.

The chemical equilibration of hadrons containing up, down and strange
quarks ($\gamma_s=1$) is very different from the situation in
$e^+e^-$, $pp$ and $p\overline p$ collisions.  In these elementary
collisions, $\gamma_s\approx0.3$; strange hadrons are suppressed by a
factor of 3 below the equilibrium expectations.

This strangeness equilibration was long-ago proposed as a signature of
the quark-gluon plasma. In a QGP, reactions proceed quickly, and
equilibrium should be reached rapidly.  Reactions are much slower in a
hadron gas and there are many species to produce, so equilibration
takes much longer.  Most (but not all) calculations indicate that the
hadron gas equilibration takes much longer than the expected system
lifetime\cite{arizona}.

%
\section{Initial States/High p$_T$ Particles}

Study of the evolution of the system before chemical freezeout
requires a probe particle that is created early in the collision.  A
few probes, such as direct photons escape the medium without
interacting, and provide information about the process that created
them.  Others, such as charmonium and high $p_T$ particles, interact
with the medium and can provide information about how it evolves.
These probes come from the hadronization of high $p_T$ quarks and
gluons (partons).  As of this conference, RHIC has so far only
presented results on the high $p_T$ hadrons.

High $p_T$ partons are produced very early in the collision, at a time
$\tau\approx\hbar/p_T$.  The partons are not expected to hadronize
until much later, around a time $t=\hbar/\Lambda$, where
$\Lambda\approx 300$ MeV is the typical QCD scale.  Usually, the
partons will have exited the medium before this point, so that
hadronization occurs in free space.  The medium will interact with the
produced parton, not the final state hadrons.  Since the final state
hadron momenta depends on the parton momentum, any energy loss by the
parton in the medium will be reflected in the high $p_T$ hadron
spectrum.

The parton energy loss can be studied by comparing hadron momentum
spectra from central heavy ion collisions with spectra from peripheral
heavy ion collisions and $pp$ collisions.  Published results, using
the 130 GeV data, have used a $pp$ reference spectrum derived from
200 GeV $p\overline p$ collision data
from the UA1 experiment.

\begin{figure}
\setlength{\epsfxsize=2.8 in}
\setlength{\epsfysize=2.3 in}
\centerline{\epsffile{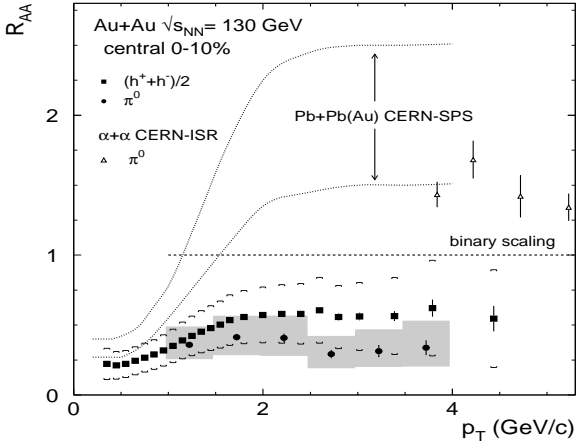}}
\caption[]{Comparison of charged hadron and $\pi^0$ $p_T$ spectra from
gold-gold collisions and a $pp$ derived reference.  From the PHENIX
collaboration\cite{PHENIXhighpt}.}
\label{phenixhighpt}
\end{figure}

Fig. \ref{phenixhighpt}\cite{PHENIXhighpt}, compares charged hadron
and $\pi^0$ production in central gold-gold collisions with a
normalized $pp$ spectrum.  $R_{AA}$ is the cross section ratio for
gold-gold to $pp$ collisions, divided by the number of nucleon-nucleon
collisions in the gold.  In the absence of nuclear effects,
$R_{AA}=1$.  At low $p_T$ soft (non-perturbative) physics is expected
to dominate, leading to $R_{AA}\ll 1$, as observed.  However, at
higher $p_T$, where perturbative QCD applies well, we expect
$R_{AA}=1$.  In contrast, at high $p_T$, in the data, $R_{AA}$
flattens out at about 0.5 for charged hadrons, and 0.3 for $\pi^0$.

Nuclear effects, such as shadowing, multiple scattering and the net
isospin difference can affect $R_{AA}$.  In $AA$ collisions, the
incident nucleons may undergo soft interactions and acquire some
initial state $p_T$ before undergoing a hard interaction.  Known as
the Cronin effect, this initial-state $p_T$ may increase the final
state $p_T$ and hence the measured R$_{AA}$.  Fig. \ref{phenixhighpt}
also shows $R_{AA}$ from lead-lead collisions at a center of mass
energy of 17 GeV/nucleon.  There, for $p_T>2$ GeV, $R_{AA}$ rises
considerably above 1.  This is dramatically different from RHIC,
showing a significant effect of the higher energy.  In fact, the lower
energy data shows no energy loss, while the RHIC data seems to
indicate a very large energy loss\cite{XNW}.

\begin{figure}
\setlength{\epsfxsize=2.8 in}
\setlength{\epsfysize=2.3 in}
\centerline{\epsffile{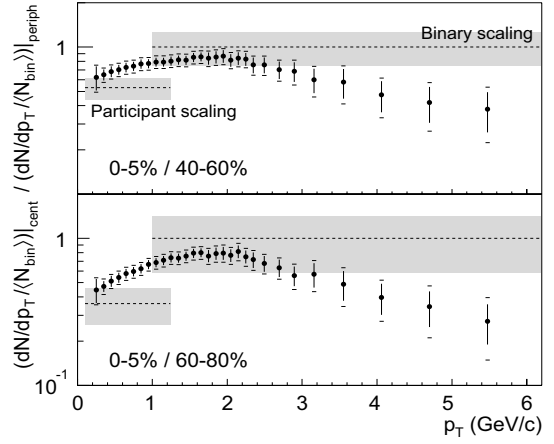}}
\caption[]{Comparison of the charged hadron $p_T$ spectra from central
and peripheral gold-gold collisions at 130 GeV. From the STAR
collaboration\cite{STARhighpt}.}
\label{starhighpt}
\end{figure}

The systematic errors in normalizing the UA1 and RHIC data vary with
the particle momentum, but are in the 35\% range, due to uncertainties
in luminosities, cross sections, centrality selection, pseudorapidity
distribution, etc.  Figure \ref{starhighpt} compares charged hadron
spectra from peripheral and central gold-gold
collisions\cite{STARhighpt}; $R_A$, the ratio of hard particle
production in central and peripheral $AA$ collisions, per
nucleon-nucleon collision, is always less than 1.  At high $p_T$,
$R_A\approx 0.3$.  The systematic uncertainties in $R_A$ are about
20\%.

\begin{figure}
\setlength{\epsfxsize=2.8 in}
\setlength{\epsfysize=2.0 in}
\centerline{\epsffile{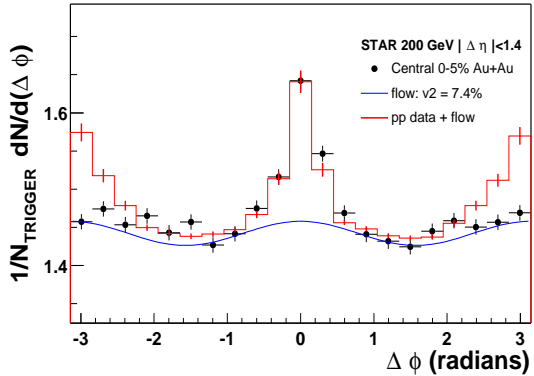}}
\caption[]{Correlation function for two high-$p_T$ particles. The
points are the 5\% most central (smallest impact parameter) 200 GeV
gold-gold collisions, while the histogram is 200 GeV $pp$ data, both
from the STAR detector\cite{Dave}.}
\label{angcorr}
\end{figure}

More can be learned about hard interactions by considering
correlations of high-$p_T$ particles.  The STAR collaboration has
studied the angular correlations between particles with $p_T>4$ GeV/c
and $|\eta|<0.7$ (the trigger particle) and a second particle with
$p_T>2$ GeV/c \cite{Dave}. Figure \ref{angcorr} shows the azimuthal
correlations, as a function of the azimuthal separation $\Delta\phi$.
There is an enhancement near $\Delta\phi=0$.  This correlation has a
similar strength and width in $pp$ and $AA$ collisions.

However, at large separations, $\Delta\phi\approx\pi$, no correlation
is observed in the $AA$ data, while a correlation is seen in the $pp$
data.  The back-to-back correlations in $pp$ collisions are expected
because jets are usually produced in back-to-back pairs known
(although many of the produced jets may be outside the experimental
acceptance).  The correlations observed in the $pp$ collisions match
the theoretical expectations, but the jet pair correlations are absent
in the $AA$ data

The major difference expected between $pp$ and $AA$ collisions is
anisotropic flow; the solid curve in Fig. \ref{angcorr} shows the size
of the flow contribution.  Flow cannot explain the difference between
the $pp$ and $AA$ curves.

The suppression of high $p_T$ particles, presence of same-side
particle correlations and disappearance of opposite side particle
correlations are all consistent with a strongly interacting system.
Only partons produced near the surface of the system are able to
escape and produce jets.  When a parton reaction produces back-to-back
partons near the system surface, one parton escapes, almost
unmodified, producing a jet.  The other parton goes in the other
direction, into the system, where it is absorbed.  The large flow at
high $p_T$ (Fig. \ref{starflow}) supports this picture, showing that
even energetic particles demonstrate collective effects.

\section{Implications for Air Shower Simulations}

This RHIC data can be used to test air shower simulations.  As several
people at the conference pointed out, the overall charged
multiplicities are considerably lower than most predictions, including
several popular air shower codes.  Models based on separate hard
(described by QCD) and soft (phenomenological) models seem to work
best.

Gluon saturation models, which can affect the depth of maximum shower
development, and the muon content of showers with energies above
$10^{17}$ eV\cite{pajares} are disfavored, but not ruled out.  In
these models, the gluon density in heavy ions becomes saturated, and
low$-x$ gluons may recombine, reducing their abundance.  They predict
that the charged particle multiplicity in $AA$ collisions should be
lower than the multiplicity scaled from $pp$ collisions; data shows
the opposite, with the $AA$ multiplicity higher than in simple $pp$
scaling.

Several other effects are likely to be relevant for air
showers. Strange particles are copiously produced, in chemical
equilibrium.  This might affect the muon content of air showers,
compared to expectations for $pp$ collisions.

The presence of non-zero net baryon density at mid-rapidity shows that
there is a substantial amount of baryon stopping, even at very high
energies.  In a fixed target frame of reference, these baryons carry
enormous energy, and so this stopping may have implications for the
overall energy flow in the collision.

Several very different analyses show that the colliding system
interacts very strongly, exhibiting collective behavior that
suppresses high $p_T$ particle production.  Pure-QCD calculations that
neglect collective effects may over-predict the number of high $p_T$
particles, and hence the shower density far from the core.

The existing RHIC data is for gold on gold collisions, not the lighter
ions and protons found in cosmic rays and the atmosphere.
Interpolation between $pp$ and gold-gold collisions is not easy.  In
the next few years, RHIC will collide lighter ions; until then, the
various Monte Carlo codes can only be tested with light (proton) or
heavy (gold) systems.

\section{Conclusions}

RHIC is just beginning it's study of ultra-relativistic heavy ion
collisions. However, already we see a few surprises: large elliptic
flow, suppression of high $p_T$ particles, and the complete chemical
equilibration of strange particles.  Chemical and thermal equilibrium
appear to have been reached.  This data shows that the system
interacts strongly and appears to equilibrate early in the collision.

But, is this the quark-gluon plasma?  The evidence suggests that the
high densities are strong interactions are consistent with a
quark-gluon plasma.  However, a very high density hadron gas cannot
yet be ruled out.

I would like to thank the STAR, PHOBOS, BRAHMS and PHENIX
collaborations for making their data and plots available to me.  This
work was funded by the U.S. Dept. of Energy under contract
DE-AC-03-76SF00098.

\end{document}